\documentclass[pre,amsmath,amssymb,showpacs,twocolumn]{revtex4}

\pdfoutput=1
\usepackage{graphicx}
\usepackage{dcolumn}
\usepackage{bm}
\usepackage{hyperref}
\usepackage{amsmath}
\usepackage[utf8]{inputenc}
\usepackage{color}

\begin{document}

\title{Diffusion and localization of relative strategy scores in the Minority Game}
\author{Mats Granath$^1$ and Alvaro Perez-Diaz $^2$}
\affiliation{$^1$Department of Physics, University of Gothenburg,
SE-41296 Gothenburg, Sweden}
\affiliation{$^2$Faculty of Engineering and the Environment, University of Southampton, SO16 7QF, UK}

\date{\today}

\begin{abstract}
We study the equilibrium distribution of relative strategy scores of agents in the asymmetric phase ($\alpha\equiv P/N\gtrsim 1$) of the basic Minority Game using sign-payoff, with $N$ agents holding two strategies over $P$ histories. We formulate a statistical model that makes use of the gauge freedom with respect to the ordering of an agent's strategies to quantify the correlation between the attendance and the distribution of strategies.  
The relative score $x\in\mathbb{Z}$ of the two strategies of an agent is described in terms of a one dimensional random walk with asymmetric jump probabilities, leading either to a static and asymmetric exponential distribution centered at $x=0$ for fickle agents or to diffusion with a positive or negative drift for frozen agents. In terms of scaled coordinates $x/\sqrt{N}$ and $t/N$ the distributions are uniquely given by $\alpha$ and in quantitative agreement with direct simulations of the game.  As the model avoids the reformulation in terms of a constrained minimization problem it can be used for arbitrary payoff functions with little calculational effort and provides a transparent and simple formulation of the dynamics of the basic Minority Game in the asymmetric phase. 


\end{abstract}

\maketitle

\section{Introduction}

A minority game can be exemplified by the following simple market analogy;  An odd number $N$ of traders (agents) must at each time step choose between two options, buying or selling a share, with the aim of picking the minority group. If sell is in minority and buy in majority one may expect the price to go up to satisfy demand and vice versa if buy is in minority, thus motivating the minority character of the game. Clearly, there is no way to make everyone content, at least half of the agents will inevitably end up in the majority group each round. As the losing agents will try to improve their lot there is no static equilibrium. 
Instead, agents might be expected to adapt their buy or sell strategies based on perceived trends in the history of outcomes \cite{farol,coop,adaptive,phase,Guinea,Cavagna1999,ChalletPRL2000,Jefferies,book,mg,review2015}. 

The Minority Game proposed by Zhang and Challet \cite{coop} formalizes this type of market dynamics where agents of limited intellect compete for a scarce resource by adapting to the aggregate input of all others \cite{farol,review2015}. Each agent has a set of strategies that, depending on the recent past history of minority groups going $m$ time steps back, gives a prediction of the next minority being buy or sell. The agent uses at each time step her highest scoring strategy which has most accurately predicted correct minority groups historically. The state space of the game is given by the strategy scores of each agent together with the recent history of minority groups, and the discrete time evolution in this space represents an intricate dynamical system. 

What makes the game appealing from a physics perspective is that it can be described using methods for the statistical physics of disordered systems, with the set of randomly assigned strategies corresponding to quenched disorder \cite{phase,impact,Continuum,ChalletPRL2000,Coolen,Coolen2005}. In particular Challet, Marsili, and co-workers showed that the model can be formulated in terms of the gradient descent dynamics of an underlying Hamiltonian \cite{Continuum}, plus noise. The asymptotic dynamics corresponds to minimizing the Hamiltonian with respect to the frequency at which agents use each strategy, a problem which in turn can be solved using the replica method \cite{replica,ChalletPRL2000,impact}. In a complementary development Coolen solved the statistical dynamics of the problem in its full complexity using generating functionals \cite{Coolen,Coolen2005,Coolen_book}.

The game is controlled by the parameter $\alpha=P/N$, where $P=2^m$ is the number of distinct histories that agents take into account, which tunes the system through a phase transition (for $N\rightarrow\infty$) at a critical value $\alpha_c=0.3374...$.
In the symmetric (or crowded) phase, $\alpha< \alpha_c$, the game is quasi-periodic with period $2P$ where a given history gives alternately one or the other of the outcomes for minority group \cite{adaptive,quasi}. A somewhat oversimplified characterization of the dynamics is that the information about the last winning minority group for a given history gives a crowding effect \cite{crowd} where many agents want to repeat the last winning outcome which then counterproductively instead puts them in the majority group. The crowding also gives large fluctuations of the size of the minority group.  

In the asymmetric (or dilute) phase, $\alpha>\alpha_c$, agents are sufficiently uncorrelated that crowding effects are not important and there is no periodic behavior. 
Instead, as exemplified in Figure \ref{timeseries} the score dynamics is random but with a net correlation between agents that makes fluctuations in the size of the minority group small. The dilute occupation of the full strategy space gives rise to a non-uniform frequency distribution of histories which can be beneficial for agents with strategies that are tuned to this asymmetry.

In this paper we study the dynamics of the Minority Game in the asymmetric phase by formulating a simplified statistical model, focusing on finding probability distributions for the relative strategy scores. In particular, we study the original formulation of the game with sign-payoff for which quantitative results are challenging to derive. By sorting the strategies based on how strongly they are correlated with the average over all strategies in the game, we find that sufficient statistical information can be extracted to formulate a quantitatively accurate model for $\alpha\gtrsim 1$.   
We discuss how the relative score for each agent can be derived from the master equation of a random walk on a chain with asymmetric jump probabilities to nearest neighbor sites, and how these jump probabilities can be calculated from the basic dynamic update equation of the scores.  The corresponding probability distributions of scores are either of the form of exponential localization or diffusion with a drift. In the appendices we show that the model is related to but independent from the Hamiltonian formulation and we show how it can also be readily applied to the game with linear payoff where the master equation has long-range hopping.\\

Although the MG is well understood from the classic works discussed above, it is our hope that the simplified model of the steady state attendance and score distributions presented in this paper provides an alternative and readily accessible perspective on this fascinating model.

\section{Definition of the Game and outline}
In order to give an overview of our results and for completeness we start by providing the formal definition of the Minority Game and some basic properties \cite{coop,book,mg}.  

At each discrete time step every agent gives a binary {\em bid} $a_i(t)=\pm 1$, all of which are collected into a total {\em attendance} 
\begin{equation}
A_t=\sum_{i=1}^Na_i(t)=-N,...,N\,,
\label{attendance}
\end{equation}
 ($N$ odd) and the winning minority group is then identified through $-\text{sign}(A_t)$. A binary string of the $m$ past winning bids, called a history $\mu$, is provided as global information to each agent upon which to base her decision for the following round. There are thus $\mu=1,...,P$ with $P=2^m$ different histories. At her disposal each agent has two randomly assigned strategies (a.k.a. strategy tables) that provide a unique bid for each history.
The bid of strategy $j=1,2$ of agent $i=1,..,N$ in response to history $\mu$ is given by $a_{i,j}^\mu=\pm 1$ and the full strategy is the $P$ dimensional random binary vector $\vec{a}_{i,j}$. There are thus a total of $2^P$ distinct strategies available. 

 The agent uses at each time step the strategy that has made the best predictions for minority group historically. This is decided by a score $U_{i,j}(t)$ for each strategy which is updated according to $U_{i,j}(t+1)=U_{i,j}(t)-a_{i,j}^{\mu}\text{sign}(A^{\mu}_t)$, irrespectively of the strategy actually being used or not. (Here the superscript $\mu$ on $A_t$ just indicates that the attendance will depend on the history $\mu(t)$ giving the bids at time $t$.) Ties, i.e. $U_{i,1}=U_{i,2}$, are decided by a coin toss. 

Since it is only the relative score between an agent's two strategies that is important in deciding which strategy to use, one may focus on the relative score 
\begin{equation}
x_i(t)=(U_{i,1}(t)-U_{i,2}(t))/2\,. 
\end{equation}
This is updated according to 
\begin{equation}
x_i(t+1)=x_i(t)+\Delta_i(t)\,,
\label{update}
\end{equation}
where
\begin{equation} 
\label{step1}
\Delta_i(t)=-\xi_i^\mu \text{sign} (A^\mu_t)\,.
\end{equation}
and where $\vec{\xi}_i=(\vec{a}_{i,1}-\vec{a}_{i,2})/2$ is an agents ``difference vector'' that takes values $\pm 1$ or $0$ for each history $\mu$.

\begin{figure}
\includegraphics[scale=.5]{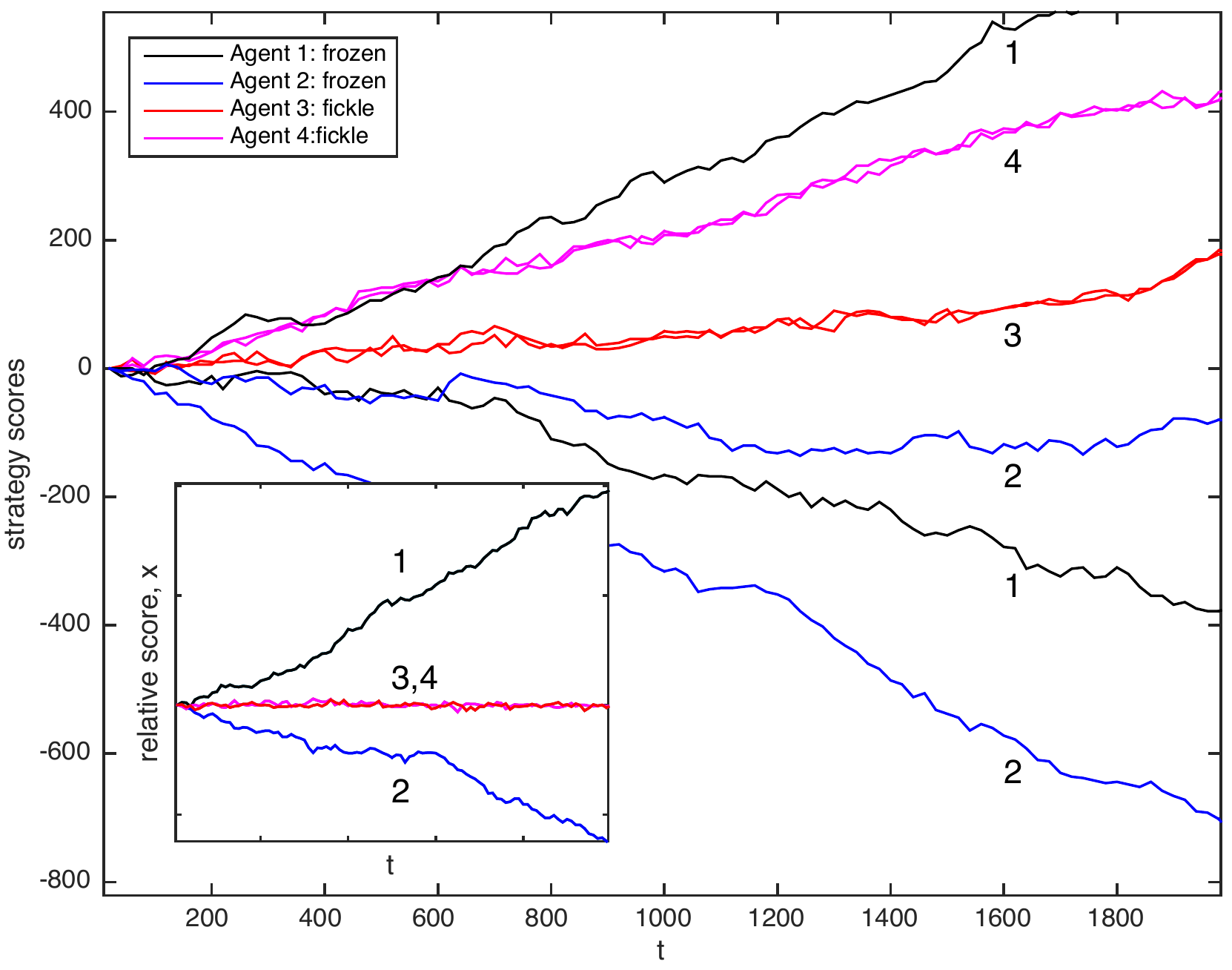}
\caption{\label{timeseries}
Evolution of strategy scores for the two strategies of four ($i=1,...4$) representative agents in a game with $N=101$ agents and a memory of length $m=7$ ($P=2^7$). At each time step every agent uses the one of her two strategies which has the highest momentary score, given by how well the strategy has predicted the past minority groups. The corresponding score difference $x_i(t)$ (inset) shows the distinction between frozen agents that consistently use a single strategy, and fickle agents that switch between strategies.
}
\end{figure}

To make the dynamics generated by these equations more concrete, Figure \ref{timeseries} shows the scores of the strategies of four particular agents $U_{i,1/2}$, $i=1,\ldots,4$ for one realization of a game with $N=101$, $P=2^7$, together with the corresponding relative scores $x_i$ (inset), over a limited time interval. As exemplified by this figure agents come in two flavors, known as ''frozen'' and ''fickle'' \cite{phase,Coolen}. An agent is frozen if one of her strategies performs consistently better than the other, such that on average the score difference is diverging, 
whereas fickle agents have a relative score that meanders around $x=0$ switching their used strategy.
The motion of $x_i$ for both fickle and frozen agents is a random walk with a bias towards or away from $x=0$.  A basic problem is to characterize and understand this random walk and derive the corresponding probability distribution $P_i(x,t)$; the probability to find agent $i$ at position $x$ at time $t$ \cite{book,Coolen_book}.

\subsection{Outline and results}
As presented in Section \ref{model} we can quantify the correlation between an agent's strategies, specified by $\xi_i^\mu$, and the total attendance $A_t^\mu$, which in turn allows for characterizing the mean (time averaged) step size $\Delta_i=\langle x_i(t+1)-x_i(t)\rangle$ in terms of a distribution over agents $P(\Delta_i)$. 
In agreement with earlier work we find that $\Delta_i$ has two contributions; one center ($x=0$) seeking bias term which arises from self interaction (the used strategy contributes to the attendance and as such is more likely to be in the majority group \cite{impact}) and a fitness term which reflects the relative adaptation of the agent's two strategies to the time averaged stochastic environment of the game. The distribution of step sizes over the population of agents are shown in Figure \ref{distribution} where frozen agents are simply those where the fitness overcomes the bias, such that $\Delta_i>0$ for $x>0$ or  $\Delta_i<0$ for $x<0$, whereas for fickle agents $\Delta_i<0$ for $x>0$ and vice versa. 


Knowing the mean step size of an agent allows for a formulation in terms of a one dimensional random walk (Fig. \ref{chain}) with corresponding jump probabilities, as presented in Section \ref{distributions}. Depending on whether it is more likely to jump towards the center or not (fickle or frozen respectively) the master equation on the chain can be solved in terms of a stationary exponential distribution centered at $x=0$ or (in the continuum limit) a normal distribution with a variance and mean that grow linearly in time (diffusion with drift). These are the distributions $P_i(x,t)$ depending on $\Delta_i$.  

In simulations over many agents 
it is natural to consider the 
full distribution $P(x,t)=\sum_{i=1}^{N} P_i(x,t)/N=\int P(\Delta_i)P_i(x,t)d\Delta_i$, with $NP(x,t)$ thus the probability of finding an agent at time $t$ with relative score $x$. In terms of scaled coordinates $x/\sqrt{N}$ and $t/N$ we find that the distribution only depends on $\alpha$.  The model distributions show excellent agreement with direct numerical simulations (Fig. \ref{Pxt} and \ref{large_a}) with no fitting parameters. This result for the full distribution of relative scores together with its systematic derivation for the original sign-payoff game represent the main results of this paper. 

In Appendix \ref{Hamiltonian} we discuss the relation between the model presented in this work and the formulation in terms of a minimization problem of a Hamiltonian generator of the asymptotic dynamics \cite{Continuum,ChalletPRL2000}. We find that one way to view the present model is as a reduced ansatz for the ground state where the only parameters are the fraction of positively and negatively frozen agents (solved for self-consistently) instead of the full space of the frequency of use of each strategy. With this ansatz closed expressions can be derived for the steady state distributions irrespective of the form of the Hamiltonian.  

In Appendix \ref{linearpay} we show how the model applies to the game with linear payoff $\Delta_i(t)=-\xi_i^\mu A^\mu_t$.

\section{Statistical model}\label{model}
We will now turn to describing the statistical model in some detail and derive the results discussed in the previous section.  
We define for each agent the sum and difference of strategies for each bid $\vec{\omega}_i=(\vec{a}_{i,1}+\vec{a}_{i,2})/2$ and (as discussed above) $\vec{\xi}_i=(\vec{a}_{i,1}-\vec{a}_{i,2})/2$ \cite{phase}. Clearly  $\omega_i^\mu$, being the sum of two random numbers $\pm 1$ is distributed over $(-1,0,1)$ with probability $(1/4,1/2,1/4)$.  A non-zero value of $\omega_i^\mu$ means that agent $i$ always has the same bid for history $\mu$ independently of which strategy it has in play. The sum over all agents, $\vec{\Omega}=\sum_{i=1}^N\vec{\omega}_i$, thus gives a constant history dependent but time independent background contribution to the attendance. (In the sense that every time history $\mu$ occurs in the time series it gives the same contribution.) This background $\Omega^\mu$ is, for large $N$, normally distributed with mean zero and variance \[\sigma_\Omega^2=N/2\,.\] 

An interesting property of the Minority Game is that there is a ``$Z_2$ gauge'' freedom with respect to an arbitrary choice of which is called strategy $1$ and which is $2$, thus corresponding to a change of sign of $\vec{\xi}_i$. Such a sign change will simply result in a change of sign of $x_i(t)$ having no consequence on which strategy is actually in play. (It is the strategy in play which is an observable, not whether it is labeled by 1 or 2.) 
Nevertheless, it turns out that making a consistent definition of the order of strategies is helpful in formulating a simple statistical model. Explicitly we order the two strategies (``fix the gauge'') of all agents $i$ such that 
\begin{equation}
\vec{\xi}_i\cdot\vec{\Omega}\leq 0\,.
\label{gauge}
\end{equation}
Shortly we will describe the distribution over agents of $\xi_i^\mu$, to quantify its anticorrelation with $\Omega_i^\mu$. 

To proceed we write the attendance at a time step $t$ with history $\mu$ as
\begin{equation}
A^\mu_t=\Omega^\mu+\sum_i\xi^\mu_i s_i(t)\,,
\label{At1}
\end{equation}
where $s_i(t)=\pm 1$ depending on which strategy agent $i$ is playing \cite{phase}. Again, the relative strategy score $x_i$ of agent $i$ is updated according to Eqn. \ref{step1}.
Given the background contribution to the attendance $\vec{\Omega}$ we expect there to be a surplus of $s_i=1$ in the steady state with our choice of gauge because the strategy 1 is expected to be favored by the score update function. (In other words, strategy 1 is expected to have a higher fitness.) However, this correlation is not trivial as the accumulated score also depends on the dynamically generated contribution the attendance.
As discussed previously some fraction $\phi$ of the agents are frozen, in the sense of always using the same strategy, $s_i=\text{constant}$. We make an additional distinction (made significant by our choice of gauge) and separate the group of frozen agents into those with $s_i(t)=1$ (fraction $\phi_1$), and those with $s_i(t)=-1$ (fraction $\phi_2$), such that $\phi=\phi_1+\phi_2$. Clearly, we expect the former to be more plentiful than the latter.    


We will now derive steady state distributions over agents for the mean step size $\Delta_i$. For this purpose we will write the attendance as 
\begin{equation}
A^{\mu}_t=\Omega^{\mu}+X^{\mu}+Y^\mu+S_t\,,
\label{Aoft}
\end{equation}
where 
\begin{eqnarray}
X^\mu&=&\sum_{i \in \phi_1}\xi^\mu_i\\
Y^\mu&=&-\sum_{i \in \phi_2}\xi^\mu_i\\
S_t &=&\sum_{i \text{ fickle}}\xi_i^\mu s_i(t)\,,
\end{eqnarray}
corresponding to the three categories of agents discussed previously. 
We will make the following simplifying approximations for these three components: the fickle component we will model as completely disordered, such that $s_i(t)=\pm 1$ is random, and correspondingly (for large N) $S_t$ is normally distributed with 
mean zero and variance \[\sigma_s^2=\varphi N/2,\] with $\varphi=(1-\phi_1-\phi_2)$ the fraction of fickle agents. (Thus, neglecting that the fickle agents would also have a net anticorrelation with the background $\vec{\Omega}$). We will assume the frozen agents to simply be a sum of independent random variables drawn from the distribution of $\vec{\xi}$, thus neglecting that the agents that are frozen may come from the extremes of this distribution. 

To proceed, we need to find the distribution of $\vec{\xi}_i$, i.e. how it varies over the set of agents. (Henceforth we will usually drop the index $i$ and regard the objects as drawn from a distribution.) Begin by defining $\vec{\psi}=\text{Random}(\pm 1)\vec{\xi}$, which is thus disordered with respect to the sign of $\vec{\Omega}\cdot\vec{\psi}$ \footnote{Note that what we here refer to as $\psi$ is what is called $\xi$ in the literature \cite{phase}. In this paper we reserve $\xi$ for the object where strategies are ordered such that $\vec{\Omega}\cdot\vec{\xi}_i\leq 0$, corresponding to $\xi_i^\mu=-\psi_i^\mu\text{sign}(\vec{\Omega}\cdot\vec{\psi}_i)$.}.   The object $\psi^\mu$ is independent of $\Omega^\mu$ (ignoring $1/N$ corrections due to $\Omega^\mu\neq 0$ limiting the available bids $\pm 1$), taking values $(1,0,-1)$ with probability $(1/4,1/2,1/4)$, which gives mean zero and variance $1/2$. Consider the joint object $h=\frac{1}{P}\vec{\Omega}\cdot\vec{\psi}$, for large $P$ this becomes normally distributed with mean zero and variance $\sigma_h^2=\frac{1}{P}(N/2)(1/2)=1/(4\alpha)$ \cite{phase}.

Now, to quantify the correlation between $\vec{\xi}$ and $\vec{\Omega}$ we define the object \[\tilde{h}=\frac{1}{P}\vec{\Omega}\cdot\vec{\xi}=-|h|\] which consequently has mean $<\tilde{h}>=-\int dhP(h)|h|=-1/\sqrt{2\pi\alpha}$ and $<\tilde{h}^2>=\sigma_h^2$. We will represent this distribution by assuming that each component $\xi^\mu$ are independent Gaussian random variables with a mean that is linearly dependent on $\Omega^\mu$. With this assumption we find the conditional distribution  
\begin{equation}
P_{\xi^\mu|\Omega^\mu}={\cal N}_{\xi^\mu}(-c(\alpha)\Omega^\mu/N,\sigma_\xi)\,,
\label{xidist}
\end{equation}
where 
$c(\alpha)=\sqrt{\frac{2}{\pi\alpha}}$, and $\sigma^2_{\xi}=1/2$, and where 
we write the normal distribution over $x$ with mean $\mu$ and variance $\sigma^2$ as ${\cal N}_x(\mu,\sigma)=\frac{1}{\sqrt{2\pi}\sigma}e^{-(x-\mu)^2/2\sigma^2}$. This quantifies that $\xi^\mu$ is on average anticorrelated with $\Omega^\mu$ which is expected to place strategy 1 in the minority group more often than strategy 2. 



Using Eqn. \ref{xidist}  we can also calculate the distributions of $X^\mu$ ($Y^\mu$) as the sum of $\phi_1 N$ ($\phi_2 N$) correlated objects $\xi_i^\mu$, giving 
\begin{eqnarray}
\label{xdist}
P_{X^\mu|\Omega^\mu}&=&{\cal N}_{X^\mu}(-c(\alpha)\phi_1\Omega^\mu,\sigma_{X|\Omega})\\
\label{ydist}
P_{Y^\mu|\Omega^\mu}&=&{\cal N}_{Y^\mu}(c(\alpha)\phi_2\Omega^\mu,\sigma_{Y|\Omega})\,,
\end{eqnarray}
with conditional variances $\sigma_{X|\Omega}^2=\phi_1N/2$ and $\sigma_{Y|\Omega}^2=\phi_2N/2$. 

\subsection{Distribution of step sizes}
Given the model expressions for the distributions of all the components of the score update equation (Eqn \ref{step1}) we will find the distribution of mean (time averaged) step sizes. 
As a first step we integrate out the fast variable $S_t$ to get a conditional on $\mu$ time averaged step size $\Delta^\mu=\langle\Delta(t)|\mu\rangle$.
(Over a long time series of the game every history $\mu$ will occur many times, we thus average over all those occurrences of a single history.)  This corresponds to
\begin{eqnarray}
\Delta^\mu&=&-\xi^\mu\int dS P(S) [\text{sign}(\Omega^\mu+X^\mu+Y^\mu+S)\nonumber\\
&+&\text{sign}(x) \xi^\mu\delta(\frac{1}{2}(\Omega^\mu+X^\mu+Y^\mu+S))]\,.
\end{eqnarray}
The second term, which is a self-interaction, follows from the discrete nature of the original problem. It gives a negative bias for the used strategy coming from the fact that if the net attendance from all other agents is zero, the used strategy puts the agent in the majority group.  (The factor $\frac{1}{2}$ in the delta function is to account for the fact that the attendance, as defined in Eqn \ref{attendance}, changes in steps of two and the factor $\text{sign}(x) \xi^\mu$ comes from the fact that only the used strategy enters the attendance.) 
Integrated this gives 
\begin{eqnarray}
\Delta^\mu&=&\Delta_{\text{fit}}^\mu+\Delta_{\text{bias}}^\mu\nonumber\\
&=&-\xi^\mu \text{erf}(\frac{\Omega^\mu+X^\mu+Y^\mu}{\sqrt{2}\sigma_S})\nonumber\\
&&-\text{sign}(x)(\xi^{\mu})^2\sqrt{\frac{2}{\pi}}\frac{1}{\sigma_S}e^{-(\Omega^\mu+X^\mu+Y^\mu)^2/2\sigma_S^2}\,,
\label{deltamu}
\end{eqnarray} 
where we have identified the first term as a fitness $\Delta_{\text{fit}}$ which quantifies the relative fitness of the agent's two strategies and the second as a negative bias $\Delta_{\text{bias}}$ for the used strategy as discussed previously. 

To calculate the distribution of mean step sizes we will assume that histories occur with the same frequency such that 
$\Delta=\frac{1}{P}\sum_\mu\Delta^\mu$. This is in fact not the case for a single realization of the game in the dilute phase, some histories occur more often than others, as one can see directly from any simulation in this regime. Nevertheless, for large $P$ we will assume that this variation of occurrences of $\mu$ averages out. As discussed extensively in the literature the overall behavior of the game is insensitive to whether the actual history is used (endogenous information) as input to the agents or if a random history is supplied (exogenous information) \cite{fake,relevance,book,mg,Coolen_book}. This is also confirmed by the present work through the good agreement between the model using exogenous information and simulations in which we use the actual history.  

Assuming large $P$ and given the assumption of independence of the distributions $\Omega,\xi,X,Y$ for different $\mu$ we expect the distribution $P(\Delta)$ to approach a Gaussian (by the central limit theorem) with mean
\begin{equation}
\bar{\Delta}=\int d\Omega d\xi dX dY P_\Omega  P_{\xi|\Omega}  P_{X|\Omega}  P_{Y|\Omega} \Delta^\mu\,,
\label{Deltaeqn}
\end{equation}
with $\Delta^\mu$ as in eqn \ref{deltamu},
 and with variance $\sigma^2=\frac{1}{P}(\overline{\Delta^2}-\bar{\Delta}^2)$.

The integrals are readily done analytically as described in the appendix \ref{app}, but the expressions are very lengthy. The main features can be expressed in the following form:
\begin{eqnarray}
\bar{\Delta}_{\text{bias}}&=&-\text{sign}(x)\frac{1}{\sqrt{N}}\tilde{\Delta}_{bias}(\alpha,\phi_1,\phi_2)\nonumber\\
\bar{\Delta}_{\text{fit}}&=&\frac{1}{\sqrt{\alpha N}}\tilde{\Delta}_{fit}(\alpha,\phi_1,\phi_2)\,,
\label{Delta_expressions}
\end{eqnarray}
where $\tilde{\Delta}_{\text{bias}/\text{fit}}>0$ are functions that only depend on $N$ and $P$ through $\alpha=P/N$, change slowly as a function of the arguments in the physically relevant regime $0\leq \phi_1+\phi_2\leq 1$ (Fig. \ref{changes}) and which satisfy $\tilde{\Delta}_{\text{bias}}(\alpha,0,0)=\frac{1}{\sqrt{2\pi}}$ and $\tilde{\Delta}_{\text{fit}}(\alpha,0,0)=\frac{1}{\pi}$. As seen from Eqn. \ref{Delta_expressions}, the mean bias is towards $x=0$, the used strategy is penalized, while the mean fitness is positive acting to increase the relative score $x$, consistent with our choice of gauge as discussed earlier. 

 The only appreciable contribution to the variance comes from the fitness term scaling as $1/P$ whereas the bias has a variance that scales with $1/(NP)$ and thus negligible (as is the cross term). The variance can be written
\begin{eqnarray}
\sigma^2_{\text{bias}}&=&0\\
\sigma^2_{\text{fit}}&=&\frac{1}{\alpha N}\tilde{\sigma}^2(\alpha,\phi_1,\phi_2)\,,
\label{sigma_expression}
\end{eqnarray}
where $\tilde{\sigma}>0$ also changes slowly in the relevant regime (Fig. \ref{changes}) and satisfies $\tilde{\sigma}(\alpha,0,0)=\frac{1}{\sqrt{6}}$. The width of the fitness distribution explains the fact that even though $\bar{\Delta}_{fit}>0$ consistent with $\phi_1\neq 0$, there are also some agents with a large negative fitness which implies $\phi_2\neq 0$. The fact that $\vec{\xi}\cdot\vec{\Omega}< 0$ thus does not necessarily imply that strategy $1$ is more successful than strategy $2$ as the correlation with the other frozen agents is also an important factor. For large $\alpha$, both the mean and variance of the fitness vanish, as can be understood as a result of there being too few agents compared to the number of possible outcomes to maintain any appreciable correlation between an agents strategies and the aggregate background, $\vec{\xi}\cdot\vec{\Omega}\approx 0$. In this limit, since the bias term always penalizes the used strategy there can be no frozen agents. We also see that both the mean and width of the distribution for given $\alpha$ scales with $1/\sqrt{N}$, consistent with simulations (Fig. \ref{distribution}).  

\subsection{Fraction of frozen agents}
\begin{figure}
\includegraphics[scale=.55]{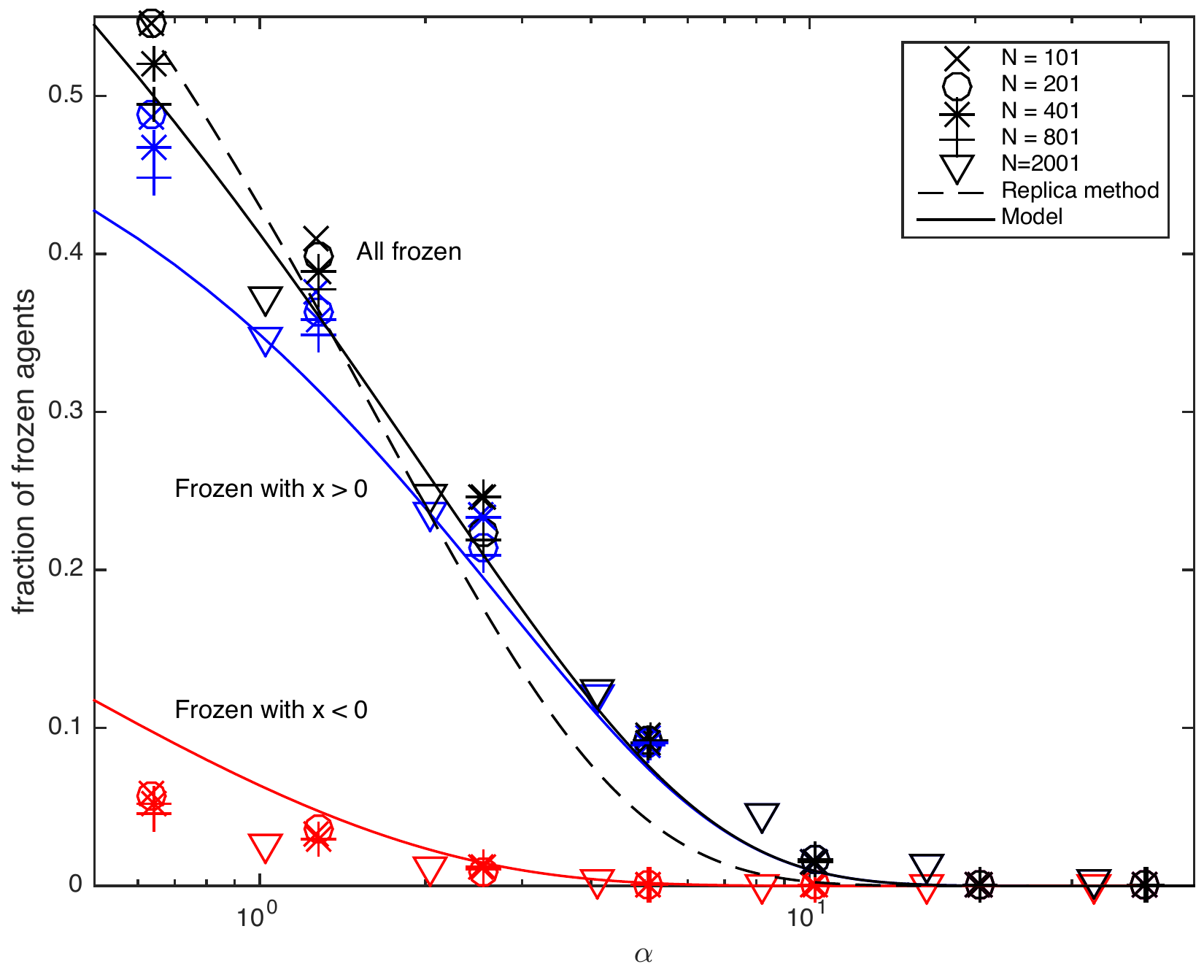}
\caption{\label{frozen_fraction}
The fraction of frozen agents as a function of $\alpha=P/N$ from the statistical model (Eqns. \ref{fraction_pos} and \ref{fraction_neg}) compared to results from direct numerical simulations of the game. The frozen agents are divided into two groups $\phi_1$ and $\phi_2$ depending on if they are frozen with relative score $x>0$ or $x<0$ respectively. The fact that $\phi_1>\phi_2$ follows from our convention $\vec{\xi}_i\cdot\vec{\Omega}\leq 0$ (eqn \ref{gauge}). Also shown is the total fraction of frozen agents from the replica calculation for linear payoff (Eqns. 3.41-3.44 of \cite{book}).   (Each data point is averaged over 20 runs with $\sim 1e6$ time steps each ($1e5$ steps for $N=2001$).)  
}
\end{figure}
\begin{figure}[h]
\includegraphics[scale=.65]{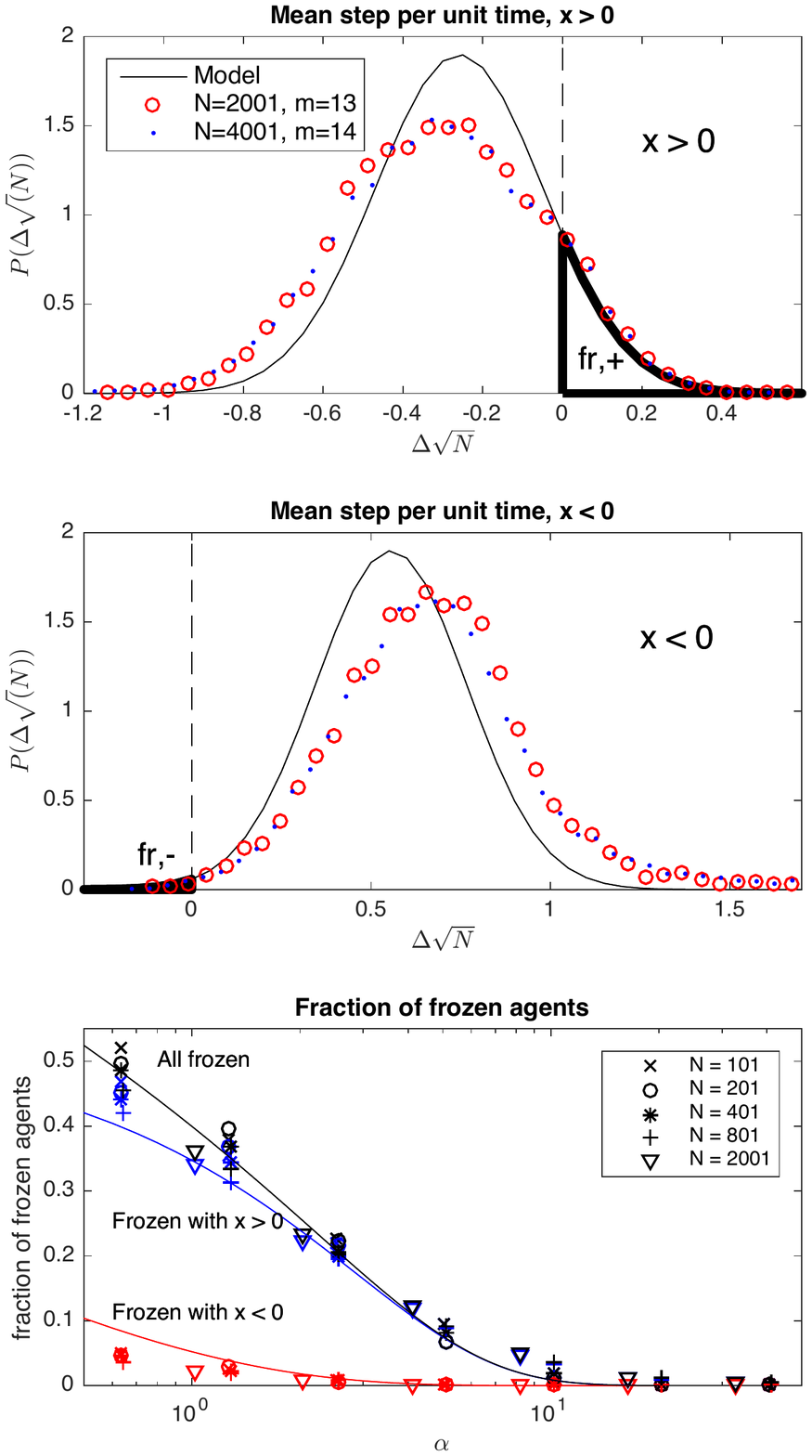}
\caption{\label{distribution}
Distributions for mean step per unit time $\Delta=\langle x(t+1)-x(t)\rangle$ at $\alpha\approx 4$ for $x>0$ (top) and $x<0$ (bottom), comparing direct simulations of the game to the statistical model (Eqn. \ref{deltaeqn}). The fraction of frozen agents with $x>0$ ($\phi_1$) is indicated by {\em ''fr,+''} and similarly for $x<0$ ($\phi_2$). The distributions of step sizes are different for $x>0$ and $x<0$ because of the convention $\vec{\xi}_i\cdot\vec{\Omega}\leq 0$ as explained in Fig. \ref{frozen_fraction}. (Simulations averaged over $1e6$ time steps, excluding a $1e4$ equilibration time.)
}
\end{figure}

For each agent the score difference $x_i$ moves with a mean step per unit time of 
\begin{eqnarray}
\Delta^+&=&\Delta_{\text{fit}}-|\bar{\Delta}_{bias}|\text{ for }x>0 \nonumber\\
\Delta^-&=&\Delta_{\text{fit}}+|\bar{\Delta}_{bias}|\text{  for }x<0\,,
\label{deltaeqn}
\end{eqnarray}
where $\Delta_{\text{fit}}$ is drawn from the distribution ${\cal N}(\bar{\Delta}_{\text{fit}},\sigma_{fit})$. If the fitness is high, such that $\Delta^+>0$, the agent will have a net positive movement and the agent is frozen, with $x_i>0$ and growing unbounded. The fraction of positive frozen agents is given by 
\begin{eqnarray}
\phi_1&=&\int_{|\bar{\Delta}_{\text{bias}}|}^\infty dz\,{\cal N}_z(\bar{\Delta}_{\text{fit}},\sigma_{\text{fit}})\nonumber\\
&&=\frac{1}{2}+\frac{1}{2}\text{erf}[\sqrt{\frac{\alpha}{2}}(\frac{\tilde{\Delta}_{\text{fit}}/\sqrt{\alpha}-|\tilde{\Delta}_{\text{bias}}|}{\tilde{\sigma}})]\,.
\label{fraction_pos}
\end{eqnarray}
Similarly, if the fitness is relatively very poor, such that $\Delta^-<0$ the agent is frozen (with $x_i<0$) with magnitude growing unbounded. The fraction of negatively frozen agents is given by 
\begin{eqnarray}
\phi_2&=&\int_{-\infty}^{-|\bar{\Delta}_{\text{bias}}|} dz\,{\cal N}_z(\bar{\Delta}_{\text{fit}},\sigma_{\text{fit}})\nonumber\\
&&=\frac{1}{2}-\frac{1}{2}\text{erf}[\sqrt{\frac{\alpha}{2}}(\frac{\tilde{\Delta}_{\text{fit}}/\sqrt{\alpha}+|\tilde{\Delta}_{\text{bias}}|}{\tilde{\sigma}})]\,,
\label{fraction_neg}
\end{eqnarray}
and correspondingly the complete fraction of frozen agents $\phi=\phi_1+\phi_2$ and fickle agents $\varphi=1-\phi$ are found. Since $\tilde{\Delta}_{\text{fit}}$, $\tilde{\Delta}_{\text{bias}}$, and $\tilde{\sigma}$ are functions of $\alpha$, $\phi_1$, and $\phi_2$, the two equations allow for solving for $\phi_1(\alpha)$ and $\phi_2(\alpha)$ as a function of the only parameter $\alpha$. We find that the solutions are readily found by forward iteration, and the results are plotted and compared to direct simulations of the game in Figure \ref{frozen_fraction} \footnote{The numerical data for Fig. \ref{frozen_fraction} is found by measuring the mean step size and identifying those that for $x>0$ have $\Delta>0$ or for $x<0$ have $\Delta<0$ as shown in Fig. \ref{distribution}.}. 
The fit is good, but there is no indication of a phase transition for small $\alpha$ in this simplified model.   

From simulations we can also measure the distribution of mean step sizes to compare to the model, which is shown in Figure \ref{distribution}. There we show an intermediate value of $\alpha$, the fit in terms of mean and width is not as good close to $\alpha_c$ and almost perfect for large $\alpha$, but everywhere the data seems well represented by a normal distribution. We also use the mean step size distributions from simulations to calculate the fraction of frozen agents, Figure \ref{frozen_fraction}. (The naive way to distinguish between frozen and switching agents; to introduce a cut-off $x_{\text{cut}}$ at some time $t$, with any agents with $|x_t|>x_{\text{cut}}$ considered frozen, makes it difficult to distinguish between frozen and switching agents with $\Delta$ near 0.)  

\section{Distributions over $x$}\label{distributions}

\begin{figure}
\includegraphics[scale=.7]{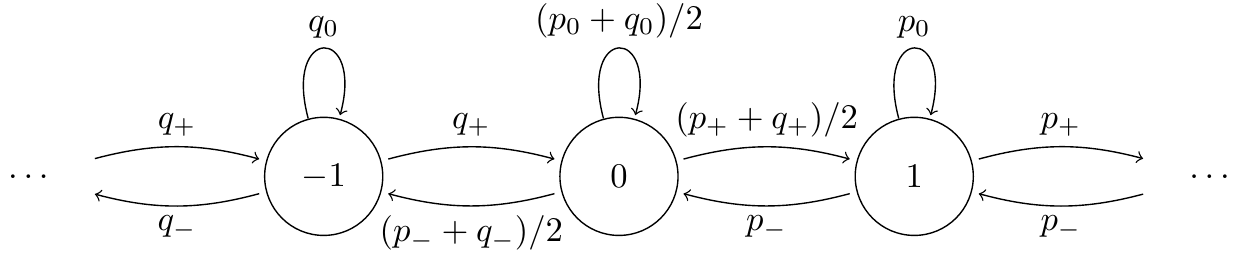}
\caption{\label{chain}
The movement of the relative strategy score $x$ of an agent is described by a random walk on a chain with jump probabilities $p_+,p_-,p_0$ for $x>1$ (i.e. strategy 1 in play) and $q_+,q_-,q_0$ for $x<-1$ (i.e. strategy 2 in play). At the boundary $x=-1,0,1$ due to the coin toss choice of strategy the probabilities are altered as in the figure.
}
\end{figure}

We now use the fact that each agent is characterized by an average step size per unit time, specified by the fitness $\Delta_{\text{fit}}$, to describe the movement of the relative score $x$ on the set of integers. Consider that the agent at time step $t$ has score difference $x$, what is the probability that at time $t+1$ the score difference is $x'$? In each time step, $x$ can only change by $-1,0,1$ as given by the basic score update equation \ref{step1}. We specify the respective probabilities $p_-,p_0,p_+$ with $p_-+p_0+p_+=1$ for $x>0$ and $q_-,q_0,q_+$ for $x<0$. 
The mean probability that $x$ remains unchanged is $p_0=q_0=\frac{1}{2}$ as this corresponds to $\xi_i^\mu=0$, meaning that the agent's two strategies have the same bid which on average (over $\mu$) will be the case for half of the histories. It should also be clear that the stepping probabilities cannot depend on the magnitude of $x$, only the sign, because the difference in score between strategies does not enter the game, only which strategy is currently used. The case $x=0$ has to be treated separately; we toss a coin to decide which strategy is used, thus the probability for a $+1$ increment is $(p_++q_+)/2$ and for a $-1$ increment is $(p_-+q_-)/2$. The movement of $x$ thus corresponds to a one-dimensional random walk on a chain, with asymmetric jump probabilities, as sketched in Figure \ref{chain}.

To relate the probabilities to the mean step size we note that for $x>0$, $\Delta^+=1\cdot p_++0\cdot p_0 -1\cdot p_-$, which together with the conservation of probability and the fact that $p_0=1/2$ gives 
\begin{eqnarray}
\label{p_eq}
p_{\pm}&=&\frac{1}{4}\pm\frac{\Delta^+}{2}\\
q_{\pm}&=&\frac{1}{4}\pm\frac{\Delta^-}{2}\,,
\end{eqnarray}
where results for $q$ follow from the same analysis for $x<0$. 
Keeping in mind that for a fickle agent $\Delta^+<0$ and $\Delta^->0$ this is of course consistent with $p_+<p_-$ and $q_-<q_+$.  A frozen agent is instead given by  $p_+>p_-$ or $q_->q_+$. 

With the known probabilities we can write down a master equation on the chain for the probability distribution $P_x(t)$ (implicit $\Delta_{\text{fit}}$ dependence)
\begin{eqnarray}
P_x(t+1)&=&p_0P_x(t)+p_+P_{x-1}(t)+p_-P_{x+1}(t),\,x>1\nonumber\\
P_x(t+1)&=&q_0P_x(t)+q_+P_{x-1}(t)+q_-P_{x+1}(t),\,x<1\,,\nonumber\\
\label{master1}
\end{eqnarray}
and at the boundary
\begin{eqnarray}
P_1(t+1)&=&p_0P_1(t)+\frac{1}{2}(p_++q_+)P_{0}(t)+p_-P_{2}(t),\nonumber\\
P_0(t+1)&=&\frac{1}{2}(q_0+p_0)P_0(t)+q_+P_{-1}(t)+p_-P_{1}(t),\nonumber\\
P_{-1}(t+1)&=&q_0P_{-1}(t)+q_+P_{-2}(t)+\frac{1}{2}(p_++q_-)P_{0}(t)\,. \nonumber\\
\label{boundary_master}
\end{eqnarray}
Assuming that the distribution is stationary, such that $P_x(t)=P_x$, and concentrating on $x>0$, we find after some manipulations the equation 
$$\frac{p_-}{p_+}-\frac{P_{x-1}}{P_x}=\frac{p_-}{p_+}\frac{P_{x+1}}{P_x}-1$$which has the exponential solution 
\begin{equation}
P_x\sim (\frac{p_-}{p_+})^{-x}=e^{-x\ln \frac{p_-}{p_+}}\approx e^{4x\Delta^+}\,,x>1\,.
\end{equation}
In the last step we used equation \ref{p_eq} and the fact that from equation \ref{Delta_expressions} the mean step size is small such that $|\Delta^+|\sim 1/\sqrt{N}\ll 1$. From this we can identify a decay length $x_+=1/(4|\Delta^+|)\sim \sqrt{N}$, which characterizes the range of positive excursions of the score difference of the fickle agent. 
Clearly, this solution requires $p_->p_+$ ($\Delta^+<0$) to be bounded, as is the case for fickle agents. 
From the same analysis for $x<1$ the fickle agents with $q_-<q_+$ have the distribution $P_x\sim e^{x\ln\frac{q_+}{q_-}}\approx e^{4x\Delta^-}$. What remains is to match up the solutions for positive and negative $x$ at the interface. This can be solved exactly, but given that the exponential prefactor is small we settle for the approximate expression 
\begin{eqnarray}
P_x&\approx&e^{-4|\Delta^+|x}P_0,\,x\geq 0\nonumber\\
P_x&\approx&e^{4\Delta^- x}P_{0},\,x\leq 0\,\nonumber\\
P_0&\approx&\frac{4}{|\Delta^+|^{-1}+(\Delta^-)^{-1}}\,.
\label{expdecay}
\end{eqnarray}
From this expression we see that the distribution is asymmetric, such that given that on average $|\Delta^+|<\Delta_-$ agents are more likely to be found with $x>0$. This opens up for a more sophisticated modelling (left for future work) where this aspect is fed back into the initial statistical description of the sum of fickle agents through the dynamical variable $S_t$, the total attendance of the fickle agents, acquiring a mean depending on $\mu$. 

For the frozen agents the master equation is the same, but given $p_+>p-$ (or $q_->q_+$) we expect a drift of the mean of the distribution. Thus focusing on long times we can consider one or the other of  Eqs. \ref{master1} depending on whether the agent is frozen with $x>0$ or $x<0$. For $x>0$ and assuming that the agent at time $t=0$ is at site $x=0$ (neglecting the influence any excursions to $x<0$)  we can write down an exact expression for $P_{x}(t)$ in terms of a multinomial distribution. Alternatively, and simpler, we can take the continuum limit $P_x(t+1)=P(x,t)+\frac{dP}{dt}$ and $P_{x\pm 1}(t)=P(x,t)\pm\frac{dP}{dx}+\frac{1}{2}\frac{d^2P}{dx^2}$ to find the Fokker-Planck equation 
\begin{equation}
\frac{\partial P}{\partial t}=-(p_+-p_-)\frac{\partial P}{\partial x}+\frac{1}{2}(p_++p_-) \frac{\partial^2 P}{\partial x^2}\,.
\end{equation}
Given the initial condition $P(x,0)=\delta(x)$ this has the solution $P(x,t)={\cal N}_x(\bar{x},\sigma_t)$ with $\bar{x}=(p_+-p_-)t=\Delta^+t$ and $\sigma^2_t=(p_++p_-)t=\frac{1}{2}t$, thus describing diffusion with a drift. 

\subsection{Full score distributions}
\begin{figure}
\includegraphics[scale=.5]{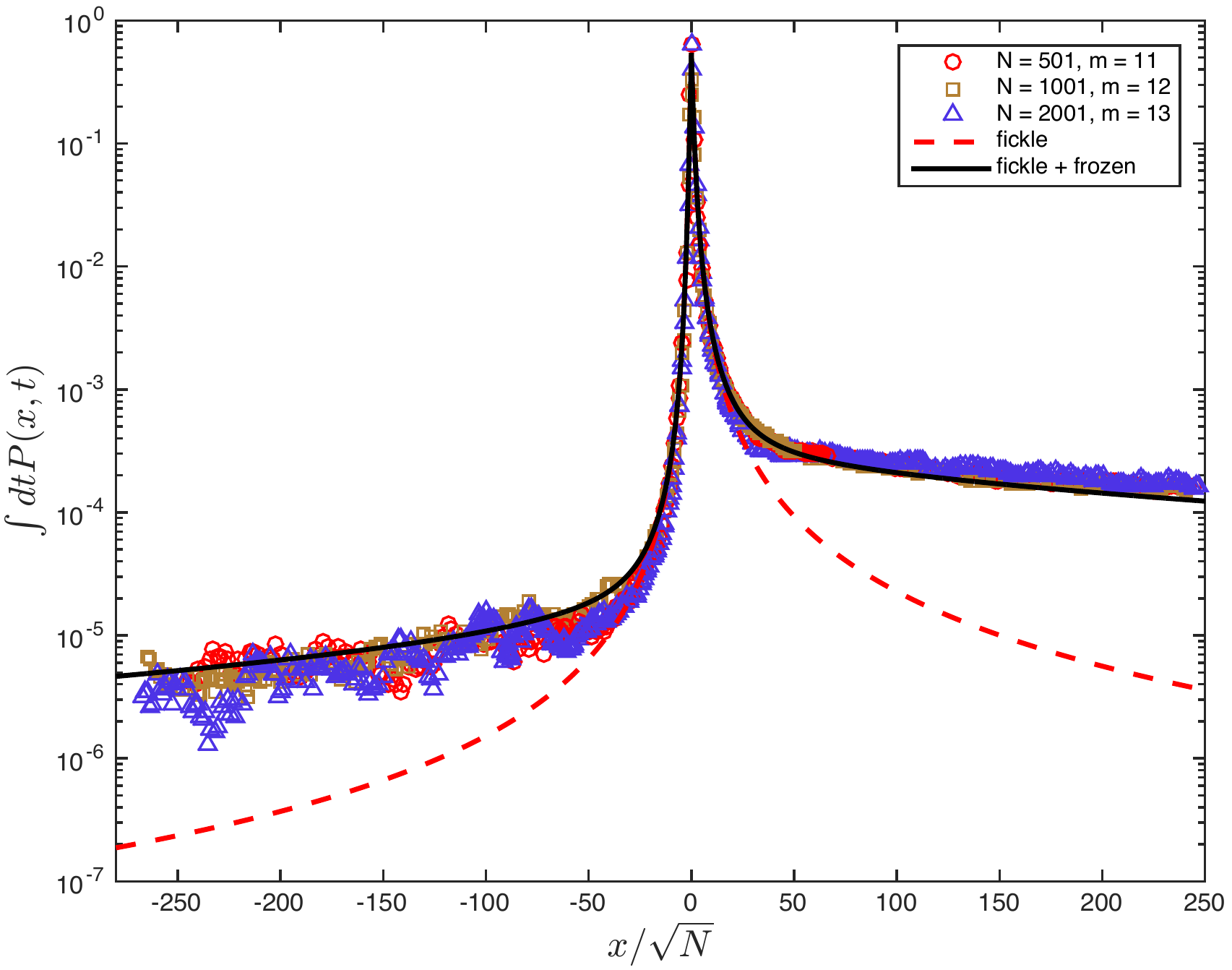}
\caption{\label{Pxt}
Full scaled distribution $P_{\tilde{x}}$ with $\tilde{x}=x/\sqrt{N}$ over all agents for $\alpha\approx 4$ compiled by averaging simulations over scaled time window $\tilde{t}_0=t_0/\sqrt{N}$ to $\tilde{t}_1=t_1/\sqrt{N}$. The model results (''fickle+frozen'') are $P_{\tilde{x}}=\frac{1}{\tilde{t}_1-\tilde{t}_0}\int_{\tilde{t}_0}^{\tilde{t}_1}d\tilde{t}P(\tilde{x},\tilde{t})$, using Equations \ref{fickle} and \ref{diffusion}.  Also shown are model results using only fickle agents. \\
The following time windows are used: for $N=501$, $t_0=5e5$ to $t_1=5e6$; for $N=1001$, $t=2t_0$ to $2t_1$; for $N=2001$,  $t=4t_0$ to $4t_1$, which correspond to the same $\tilde{t}_0$ and $\tilde{t}_1$. (Simulations are averaged over 80 runs for $N=501$ and 15 runs for $N=1001$ and $2001$.) 
}
\end{figure}

\begin{figure}
\includegraphics[scale=.5]{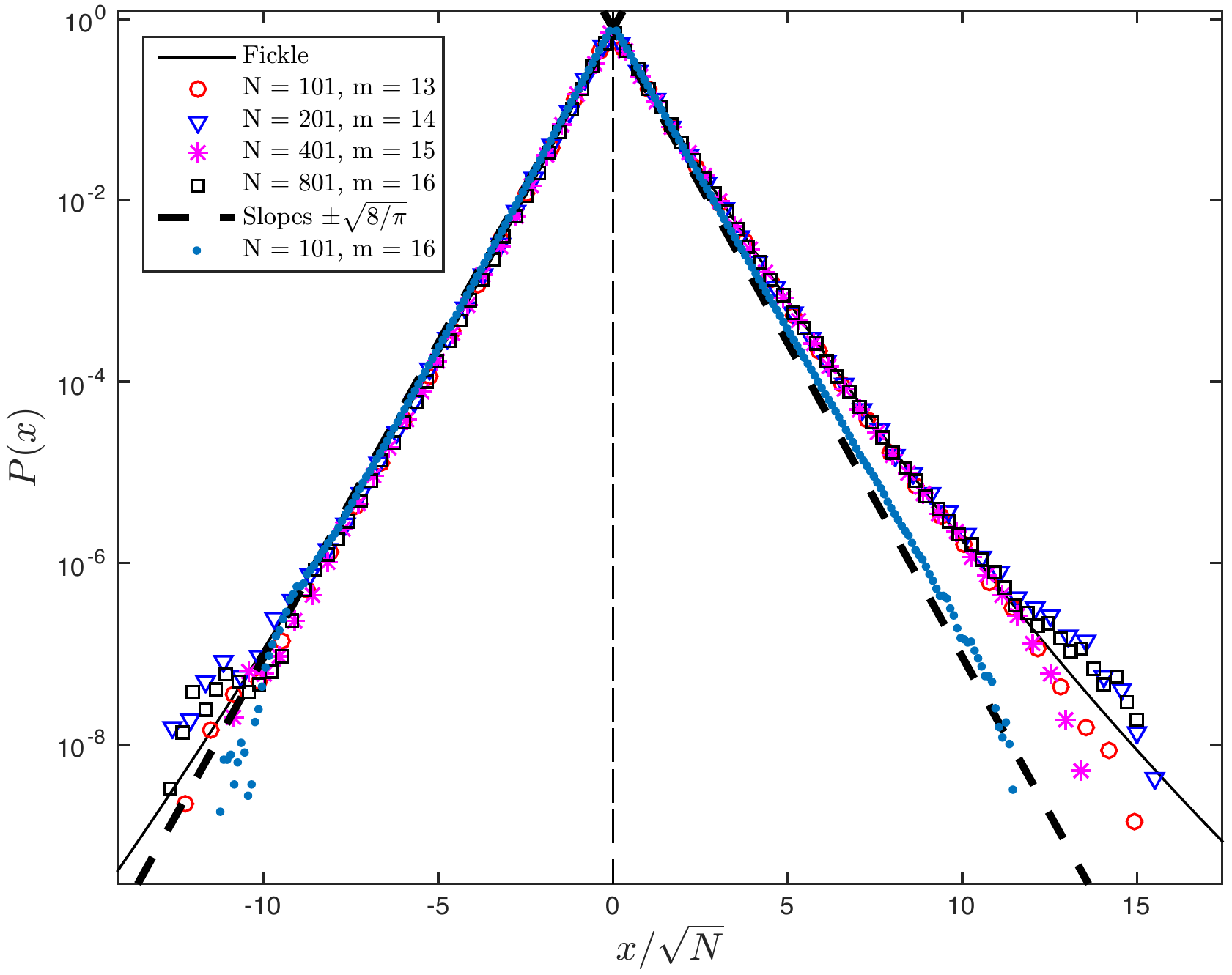}
\caption{\label{large_a}
Distribution $P_{\tilde{x}}$ at large $\alpha\approx 80$. There are no frozen agents, and the simulated and model (``fickle'') distributions are stationary. Also shown is the asymptotic $\alpha\rightarrow\infty$ behavior where all agents are symmetrically localized with localization length $x_0=\sqrt{\pi N/8}$, and a simulation at $\alpha\approx 650$ which approaches this asymptotic behavior.  (Simulations averaged over $\sim 4e8$ time steps.)
}
\end{figure}

Given that we now have a description of the relative score distribution of a single agent in terms of an asymmetric exponential decay or diffusion, we can also consider the full distribution of relative scores over all agents, by integrating over the distribution of mean step sizes. Defining the scaled variables $\tilde{x}=x/\sqrt{N}$ and $\tilde{t}=t/N$ we 
write $P(\tilde{x},\tilde{t})=P_{\text{fi}}(\tilde{x})+P_{\text{fr},+}(\tilde{x},\tilde{t})+P_{\text{fr},-} (\tilde{x},\tilde{t})$, corresponding to the stationary distribution of the fickle agents and diffusive distributions of the frozen agents with $x>0$ and $x<0$ respectively. The first component is   
\begin{equation}
P_{\text{fi}}(\tilde{x})=\int_{-b_\alpha}^{b_\alpha}\frac{dz\, {\cal N}_z(\frac{\tilde{\Delta}_{\text{fit}}}{\sqrt{\alpha}},\frac{\tilde{\sigma}}{\sqrt{\alpha}})\,4e^{4(z\pm b_\alpha)\tilde{x}}}{(b_\alpha-z)^{-1}+{(b_\alpha+z)^{-1}}}\,,
\label{fickle}
\end{equation}
where $\pm$ corresponds to $x<0$ and $x>0$ respectively, and where $b_\alpha=|\tilde{\Delta}_{\text{bias}}|$.
For the frozen agents we have 
\begin{eqnarray}
P_{\text{fr},+}(\tilde{x},\tilde{t})&=&\int_{b_\alpha}^{\infty}dz\, {\cal N}_z(\frac{\tilde{\Delta}_{\text{fit}}}{\sqrt{\alpha}},\frac{\tilde{\sigma}}{\sqrt{\alpha}})\,  {\cal N}_{\tilde{x}}(\tilde{t}(z-b_\alpha),\sigma_{\tilde{t}})\nonumber\\
P_{\text{fr},-}(\tilde{x},\tilde{t})&=&\int_{-\infty}^{-b_\alpha}dz\, {\cal N}_z(\frac{\tilde{\Delta}_{\text{fit}}}{\sqrt{\alpha}},\frac{\tilde{\sigma}}{\sqrt{\alpha}})\,  {\cal N}_{\tilde{x}}(\tilde{t}(z+b_\alpha),\sigma_{\tilde{t}})\,,\nonumber\\
\label{diffusion}
\end{eqnarray}
where $\sigma^2_{\tilde{t}}=\tilde{t}/2$.
These expressions are compared to direct simulations of the game  for intermediate $\alpha\approx 4$ in Fig. \ref{Pxt}. The simulations are averaged over a specific time window and the diffusive component Eqn. \ref{diffusion} is integrated over the corresponding scaled time window. The agreement is excellent over the complete stationary and diffusive components of the distribution and shows the data collapse in terms of scaled coordinates. In Fig. \ref{large_a} we also show a comparison for large $\alpha\approx 80$ where the simulations have no frozen agents and all fickle agents are localized by a length close to the $\alpha\rightarrow\infty$ value $x_0=\sqrt{\pi N/8}$.

The asymmetry of these plots is an artefact of our gauge choice $\vec{\xi}_i\cdot\vec{\Omega}\leq 0$ which implies that on average agents will use strategy 1 ($x>0$) more frequently than strategy 2 ($x<0$). To restore the full symmetry is simply a matter of symmetrizing the distributions around $x=0$. 

Finally, we remark that the formal solution in terms of an exponential distribution of strategy scores for frozen agents was derived in \cite{Continuum} from a Fokker-Planck equation for the linear payoff game. See Appendix \ref{Hamiltonian} and \ref{linearpay} for a further discussion of the comparison between the present model and the Hamiltonian formulation.

\section{Summary}

We have studied the asymmetric phase of the basic Minority Game, focusing on the statistical distribution of relative strategy scores and the original sign-payoff formulation of the game. 
We formulate a  statistical model for the attendance that relies on a specific gauge choice in which the two strategies of each agent are ordered with respect to the background ($\vec{\xi}_i\cdot\vec{\Omega}\leq 0$ for all agents $i$). Using this model we can derive a distribution of the mean step per time increment for the relative scores, specified in terms of a bias for the used strategy and the relative fitness of the two strategies. The relative strategy score for each agent is conveniently described as a random walk on an integer chain, where the jump probabilities are calculated from the mean step. The probability distribution of observing the agent at some position on the chain at a given time is either given by a static asymmetric exponential localized around $x=0$ for fickle agents or to diffusion with a drift for frozen agents. Excellent agreement with direct simulations of the game for the score distribution confirms the basic validity of the modelling. At the same time, as discussed in the appendix, the fluctuations of the attendance are overestimated by the model. By contrasting with the Hamiltonian formulation of the dynamics the reason for this discrepancy is readily understood from viewing the model as a crude ansatz for full minimization problem. This also opens up for improving the model by introducing some variational parameters without having to confront the full complexity of the minimization of a non-quadratic Hamiltonian for general payoff functions. \\

 

We thank Erik Werner for valuable discussions. Simulations were performed on resources at Chalmers Centre for Computational Science and Engineering (C3SE) provided by the Swedish National Infrastructure for Computing (SNIC).

\appendix
\section{Solving for mean and variance of step size. \label{app}}
\begin{figure}[h]
\includegraphics[scale=.5]{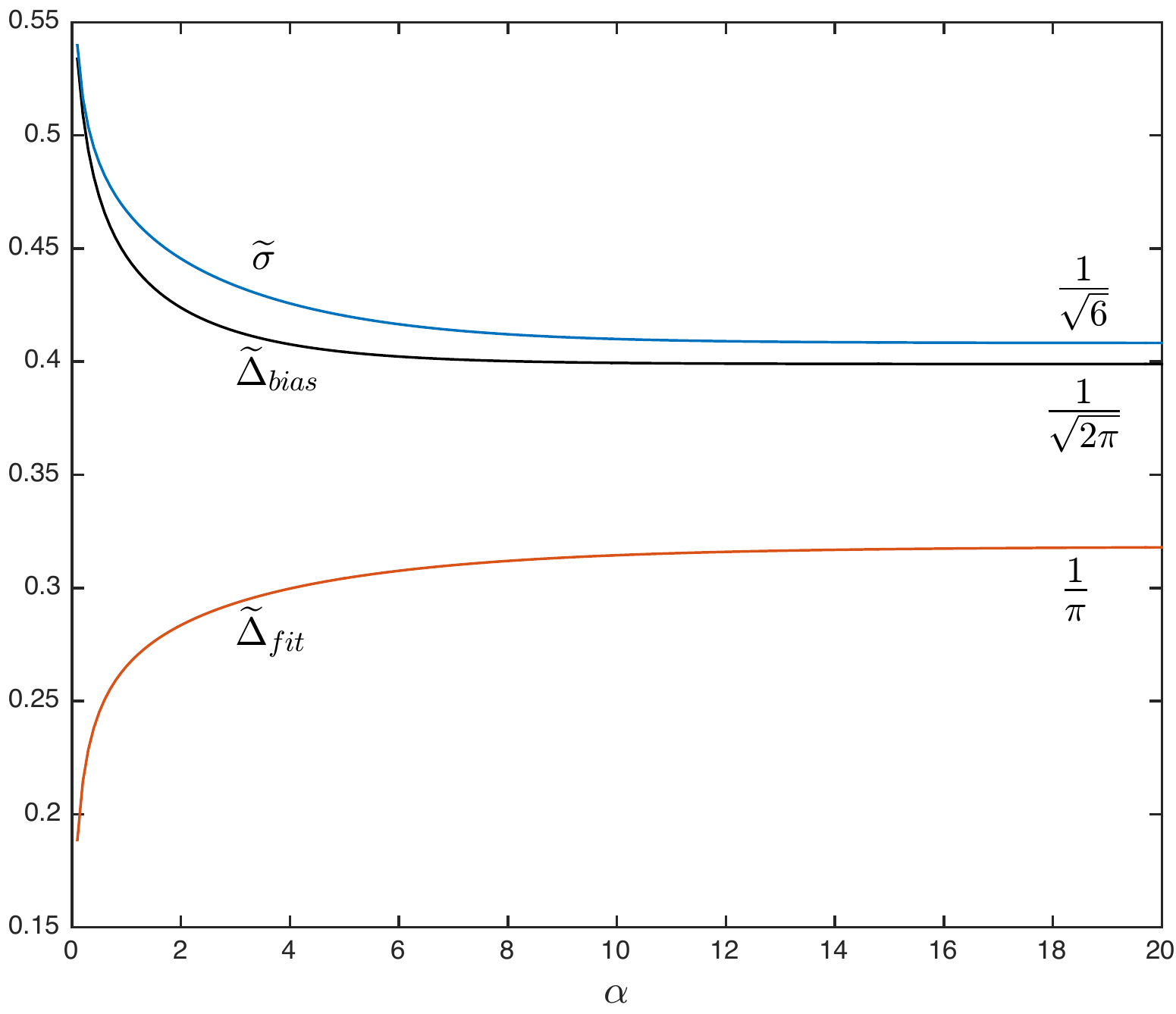}
\caption{\label{changes}
The $\alpha$ parameter dependence of the three quantities specifying the mean and variance of the distribution of mean step sizes according to Equations \ref{Delta_expressions} and \ref{sigma_expression}. 
}
\end{figure}

The integrals to calculate the mean and variance for the distribution of average step sizes, Eqn. \ref{Deltaeqn}, are Gaussian integrals including the error function. To solve these we first rescale the variables in terms of the variance $\Omega/\sigma_\Omega\rightarrow\Omega$, $X/\sigma_{X|\Omega}\rightarrow X$ etc. and perform the integral over the distribution of agents $\xi$ which evaluates to $\langle \xi|\Omega \rangle=-c(\alpha)\Omega/\sqrt{2 N}$ ($c(\alpha)=\sqrt{\frac{2}{\pi\alpha}}$) and $\langle \xi^2|\Omega \rangle=\frac{1}{2}$. We are left with integrals 
\begin{eqnarray}
\bar{\Delta}_{\text{bias}}&=&-\text{sign}(x)\frac{1}{\sqrt{\pi N\varphi}}\int\frac{d\Omega dXdY}{(2\pi)^{3/2}}\nonumber\\
&&e^{-\frac{1}{2}[\Omega^2+(X+\sqrt{\phi_1}c(\alpha)\Omega)^2+(Y-\sqrt{\phi_2}c(\alpha)\Omega)^2]}\nonumber\\ &&e^{-\frac{1}{2}(\frac{\Omega+\sqrt{\phi_1}X+\sqrt{\phi_2}Y}{\sqrt{\varphi}})^2}\,,
\end{eqnarray}

\begin{eqnarray}
\bar{\Delta}_{\text{fit}}&=&\frac{c(\alpha)}{\sqrt{2 N}}\int\frac{d\Omega dXdY}{(2\pi)^{3/2}}\nonumber\\
&&e^{-\frac{1}{2}[\Omega^2+(X+\sqrt{\phi_1}c(\alpha)\Omega)^2+(Y-\sqrt{\phi_2}c(\alpha)\Omega)^2]}\nonumber\\ &&
\Omega\, \text{erf}(\frac{\Omega+\sqrt{\phi_1}X+\sqrt{\phi_2}Y}{\sqrt{2\varphi}})\,,
\end{eqnarray}
and 
\begin{eqnarray}
\sigma^2_{\text{fit}}&=&\frac{1}{2 P}\int\frac{d\Omega dXdY}{(2\pi)^{3/2}}\nonumber\\
&&e^{-\frac{1}{2}[\Omega^2+(X+\sqrt{\phi_1}c(\alpha)\Omega)^2+(Y-\sqrt{\phi_2}c(\alpha)\Omega)^2]}\nonumber\\ &&
\text{erf}^2(\frac{\Omega+\sqrt{\phi_1}X+\sqrt{\phi_2}Y}{\sqrt{2\varphi}})\,,
\end{eqnarray}
To evaluate these we use the following integral formulas  
\begin{equation}
\int(\prod_i \frac{d x_i}{\sqrt{2\pi}})e^{-\frac{1}{2}x^TAx}=1/\sqrt{det(A)} \,,
\end{equation}
\begin{equation}
\int\frac{dx}{\sqrt{2\pi}}e^{-\frac{x^2}{2}}x\,\text{erf}(\frac{bx}{\sqrt{2}}+c)=\sqrt{\frac{2}{\pi}}\frac{b}{\sqrt{1+b^2}}e^{-c^2/(1+b^2)} \,,
\end{equation}
and
\begin{equation}
\int\frac{dx}{\sqrt{2\pi}}e^{-\frac{x^2}{2}}\text{erf}^2(\frac{bx}{\sqrt{2}})=\frac{4}{\pi}\arctan\sqrt{1+2b^2}-1\,,
\end{equation}
where $A$ is a symmetric (positive definite) matrix, and $b$ and $c$ are real constants. 
The bias term thus follows from a direct application of the first integral formula to a 3x3 matrix. The fitness term follows from a substitution $X'=X+\sqrt{\phi_1}c(\alpha)\Omega$ and $Y'=Y-\sqrt{\phi_2}c(\alpha)\Omega$ to apply the second integral formula over $\Omega$ and subsequently the first integral formula on a 2x2 matrix. The variance can be calculated by the substitution for $\Omega$, $z=\Omega+\sqrt{\phi_1}X+\sqrt{\phi_2}Y$, followed by integrating out $X$ and $Y$ to finally apply the third integral formula over $z$. 
The actual expressions are quite lengthy\footnote{The exact expressions for these quantities are derived from the integral formulas as explained, but we are also happy to share them directly. Contact the first author.}, but the important features can be represented according to Eqs. \ref{Delta_expressions} and \ref{sigma_expression} in terms of functions 
$\tilde{\Delta}_{\text{bias}}(\alpha,\phi_1,\phi_2)$,  $\tilde{\Delta}_{\text{fit}}(\alpha,\phi_1,\phi_2)$, and $\tilde{\sigma}(\alpha,\phi_1,\phi_2)$. After solving for for the fractions of frozen agents 
$\phi_1(\alpha)$ and $\phi_2(\alpha)$ using Eqs. \ref{fraction_pos} and \ref{fraction_neg}, we can consider these functions as dependent only on the control parameter $\alpha$. The dependence on $\alpha$ is plotted in Figure \ref{changes}, to point out that these functions change little over the whole relevant range $\alpha>\alpha_c\approx 0.3$.  

\section{Hamiltonian formulation\label{Hamiltonian}}
Here we connect the formalism in the present work to the solution using the replica method, following closely the presentation in \cite{Continuum} and \cite{ChalletPRL2000}. 
Expressing the attendance for given history in terms of fluctuations around a mean as  
\begin{equation}
A_t^\mu=\langle A |\mu\rangle +S_t\,,
\label{At}
\end{equation}
where $S_t$ is a Gaussian random variable with mean zero and variance $\sigma_S^2$ (to be determined self-consistently). This is related to expression (\ref{Aoft}), where we take an explicit statistical form $\langle A|\mu\rangle=\Omega^\mu+X^\mu+Y^\mu$, assumed to correspond to background plus frozen agents. Also, in the model in this paper we have the magnitude of $\sigma_S^2$ as $\varphi N/2$, with $\varphi$ the fraction of fickle agents. This is not assumed in the present treatise, but as we will see the outcome is related.  

There is also the explicit expression, Eqn. \ref{At1}, for the attendance
$A_t^\mu=\Omega^\mu+\sum_i\xi_i^\mu s_i(t)$,
where $s_i(t)=\pm 1$ depending on which strategy is momentarily used by the agent.  Taking the time average of this and assuming that the frequency of use is not influenced by the rapid switches of history we write $\langle s_i(t)\rangle=m_i$, where for frozen agents $m_i=\pm 1$ and for fickle $|m_i|<1$. AS discussed in \cite{Continuum} the fluctuations of $s_i(t)$ are statistically independent such that $\langle s_i(t)s_j(t)\rangle=m_im_j$ for $i\neq j$, whereas $(s_i(t))^2=1$ by definition. 
With this we can write $\langle A|\mu\rangle=\Omega^\mu+\sum_i\xi_i^\mu m_i$, noting that $\frac{\partial  \langle A|\mu\rangle}{\partial m_i}=\xi_i^\mu$. 

Now, evaluating the variance of the attendance using Eqn. \ref{At1} and $\sigma^2_\Omega =N/2$, we find \[\sigma^2=\langle A^2\rangle=\frac{N}{2}+\frac{1}{P}(2\vec{\Omega}\cdot\vec{\xi_i} m_i+\sum_{i\neq j}\vec{\xi_i}\cdot\vec{\xi_j}m_im_j+\sum_i(\vec{\xi}_i)^2)\,.\] This we can  alternatively write (using Eqn. \ref{At}) as $\sigma^2=\frac{1}{P}\sum_\mu\langle A |\mu\rangle^2+\sigma_S^2=H+\sigma_S^2$. Here $H$, the {\em predictability}, also has the alternative form (using Eqn. \ref{At1}) \[H=\frac{N}{2}+ \frac{1}{P}(2\vec{\Omega}\cdot\vec{\xi_i} m_i+\sum_{ij}\vec{\xi_i}\cdot\vec{\xi_j}m_im_j)\,.\] Correspondingly we find for the rapidly fluctuating field $S_t$ the variance \[\sigma_S^2=\sigma^2-H=\sum_i\frac{1}{2}(1-m_i^2)\] (using $\sigma_{\xi}^2=1/2$). The latter expression has no contribution from frozen agents (as expected), and assuming that the distribution of $m_i$ is quite strongly centred at $0$ it will be close to, but always lower than, our assumed value of $\varphi N/2$.

Consider now the fixed history time averaged step size for agent $i$, $\Delta_i^\mu=-\xi_i^\mu \langle \text{sign}(A_t)|\mu\rangle$, with 
\[\langle \text{sign}(A_t)|\mu\rangle=\int dSP(S)\text{sign}(\langle A |\mu\rangle+S).\]
The aim is to find a Hamiltonian generator ${\cal H}$ of the long time dynamics such that the time and history averaged update is given by 
\[\bar{\Delta}_i=\frac{1}{P}\sum_\mu \xi_i^\mu\langle \text{sign}(A)|\mu\rangle=-\frac{\partial {\cal H}}{\partial m_i}.\]
(Note that this expression is not equivalent to Eqn. \ref{Deltaeqn}). The latter is the mean of a distribution, whereas the present object represents the full distribution of average step sizes over agents corresponding to different $i$.) A function that does this is ${\cal H}=\int dS P(S) G(\langle A |\mu\rangle+S)$ where $G(x)=x\,\text{sign}(x)$ such that $\frac{dG}{dx}=\text{sign}(x)$, which evaluates to
\begin{equation}
{\cal H}=\frac{1}{P}\sum_\mu (\langle A |\mu\rangle \text{erf}(\frac{\langle A |\mu\rangle}{\sqrt{2}\sigma_S})+\sqrt{\frac{2}{\pi}}\sigma_Se^{-(\langle A |\mu\rangle)^2/2\sigma_s^2})\,.
\label{hcal}
\end{equation}

Thinking of the long-time evolution of the score difference for agent $x_i$ which has an average step size $\bar{\Delta}_i$, we find that if $\bar{\Delta}_i>0$ the agent will be frozen positive, with $m_i=1$ and similarly  if $\bar{\Delta}_i<0$ it will be frozen negative, with $m_i=-1$. Only if $\bar{\Delta}_i=0$ the agent will be fickle, with $-1<m_i<1$. Considering that $\bar{\Delta}_i=-\frac{\partial {\cal H}}{\partial m_i}$ we find the three cases: $m_1=1$ corresponds to $\frac{\partial {\cal H}}{\partial m_i}<0$, $m_1=-1$ corresponds to $\frac{\partial {\cal H}}{\partial m_i}>0$, and $-1<m_i<1$ corresponds to  $\frac{\partial {\cal H}}{\partial m_i}=0$. The solution to this thus corresponds to finding the minimum of ${\cal H}$ with respect to $\{m_i\}$. 

The minimization of Eqn. \ref{hcal} however, looks like a formidable problem in the thermodynamic limit, and we are not aware that it has been pursued in the literature. (Note that $\langle A|\mu\rangle\sim\sqrt{N}\sim\sigma_S$ such that an expansion is not appropriate.) This is in contrast to the case of linear payoff (se Appendix \ref{linearpay}) where 
${\cal H}_{\text{linear}}=H=\frac{1}{P}\sum_\mu \langle A|\mu\rangle^2$ which is a quadratic form in the variables $m_i$. For the latter case  the minimization problem has been solved using the replica method \cite{replica,ChalletPRL2000,impact}. The equilibrium score distributions that we focus on in the present work have been solved for in \cite{Continuum} but to the best of our knowledge not for the sign-payoff game. Also, it appears that these distributions have not been discussed or studied in any detail, or compared to simulations, in earlier work.

\section{Distributions with linear payoff \label{linearpay}}
\begin{figure}[h]
\includegraphics[scale=.5]{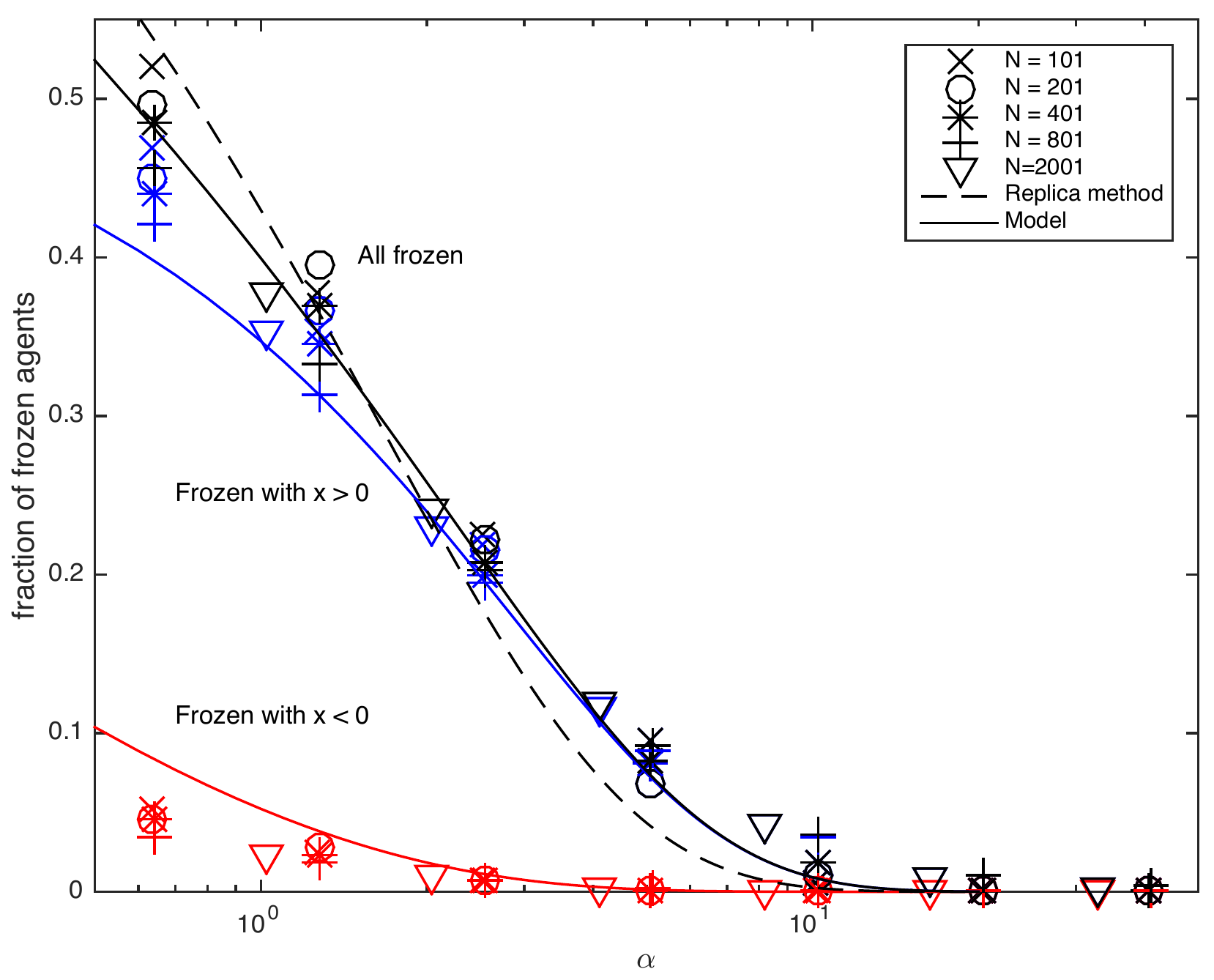}
\caption{\label{linear_fracs}
The fraction of frozen agents as a function of $\alpha$ for linear payoff. Also shown is the total fraction of frozen agents from the replica calculation (Eqns. 3.41-3.44 of \cite{book}) (Each data point is averaged over 20 runs with $\sim 1e6$ time steps each ($1e5$ steps for $N=2001$).)  }
\end{figure}

Here we repeat the analysis of the main paper for the case of linear payoff where Eqn. \ref{step1} is replaced by 
\begin{equation} 
\label{steplin}
\Delta_i(t)=-\xi_i^\mu A^\mu_t\,.
\end{equation}
We apply the same distributions, Eqs. \ref{xidist}-\ref{ydist}, for the relative bid $\xi^\mu$, the contribution to the attendance of the positively ($x>0$) frozen agents $X^\mu$, and the negatively ($x<0$) frozen agents $Y^\mu$ and
write $A_t^\mu=\Omega^{\mu}+X^{\mu}+Y^\mu+S_t$ (Eqn. \ref{Aoft}). Here $\Omega^\mu$ is the background (mean zero, variance $N/2$) and $S_t$ is the contribution from the fickle agents (with assumed mean zero). Integrating over time at fixed history $\mu$,   
$S_t$ integrates to zero because of linearity, giving
\begin{eqnarray}
\Delta^\mu&=&\Delta_{\text{fit}}^\mu+\Delta_{\text{bias}}^\mu\nonumber\\
&=&-\xi^\mu (\Omega^\mu+X^\mu+Y^\mu)-\text{sign}(x) (\xi^\mu)^2\,,
\end{eqnarray} 
where we have explicitly inserted the negative bias term for the used strategy. Averaging over histories in the large $P$ limit we find that the bias is just a constant 
\begin{equation}
\Delta_{\text{bias}}=-\text{sign}(x)\frac{1}{2}\,,
\end{equation}
and the fitness is normal with mean and variance given by 
\begin{eqnarray}
\bar{\Delta}_{\text{fit}}&=&\frac{c}{2}(1-c\phi_1+c\phi_2)\\
\sigma^2_{\text{fit}}&=&\frac{1}{4\alpha}[(1-c\phi_1+c\phi_2)^2+\phi_1+\phi_2]\,,
\end{eqnarray}
where as before $c=c(\alpha)=\sqrt{2/\pi\alpha}$ and $\phi_1$ and $\phi_2$ are the respective fractions of frozen agents. We note that the step size is of order $1$ for the linear payoff, compared to order $1/\sqrt{N}$ for the sign payoff game. Similarly in both cases, for large $\alpha$ the fitness drops out, ensuring that there are no frozen agents.  For moderate $\alpha$ the fraction of frozen agents need to be solved for self-consistently through the equations
\begin{eqnarray}
\phi_1&=&\int_{1/2}^{\infty}dz{\cal N}_z(\bar{\Delta}_{\text{fit}},\sigma_{\text{fit}})=\frac{1}{2}\text{erfc}(\frac{\frac{1}{2}-\bar{\Delta}_{\text{fit}}}{\sqrt{2}\sigma_{\text{fit}}})\nonumber\\
\phi_2&=&\int_{-\infty}^{-1/2}dz{\cal N}_z(\bar{\Delta}_{\text{fit}},\sigma_{\text{fit}})=\frac{1}{2}\text{erfc}(\frac{\frac{1}{2}+\bar{\Delta}_{\text{fit}}}{\sqrt{2}\sigma_{\text{fit}}})\,.\nonumber
\end{eqnarray}
As for the sign-payoff game the results from solving these equations numerically are in good agreement with simulation data in the dilute phase as shown in Fig. \ref{linear_fracs}. (Note, compared to Fig. \ref{frozen_fraction}, that both the data and model results for the fraction of frozen agents are very similar and quite insensitive to whether sign-payoff or linear payoff is used.) 

\begin{figure}[h]
\includegraphics[scale=.5]{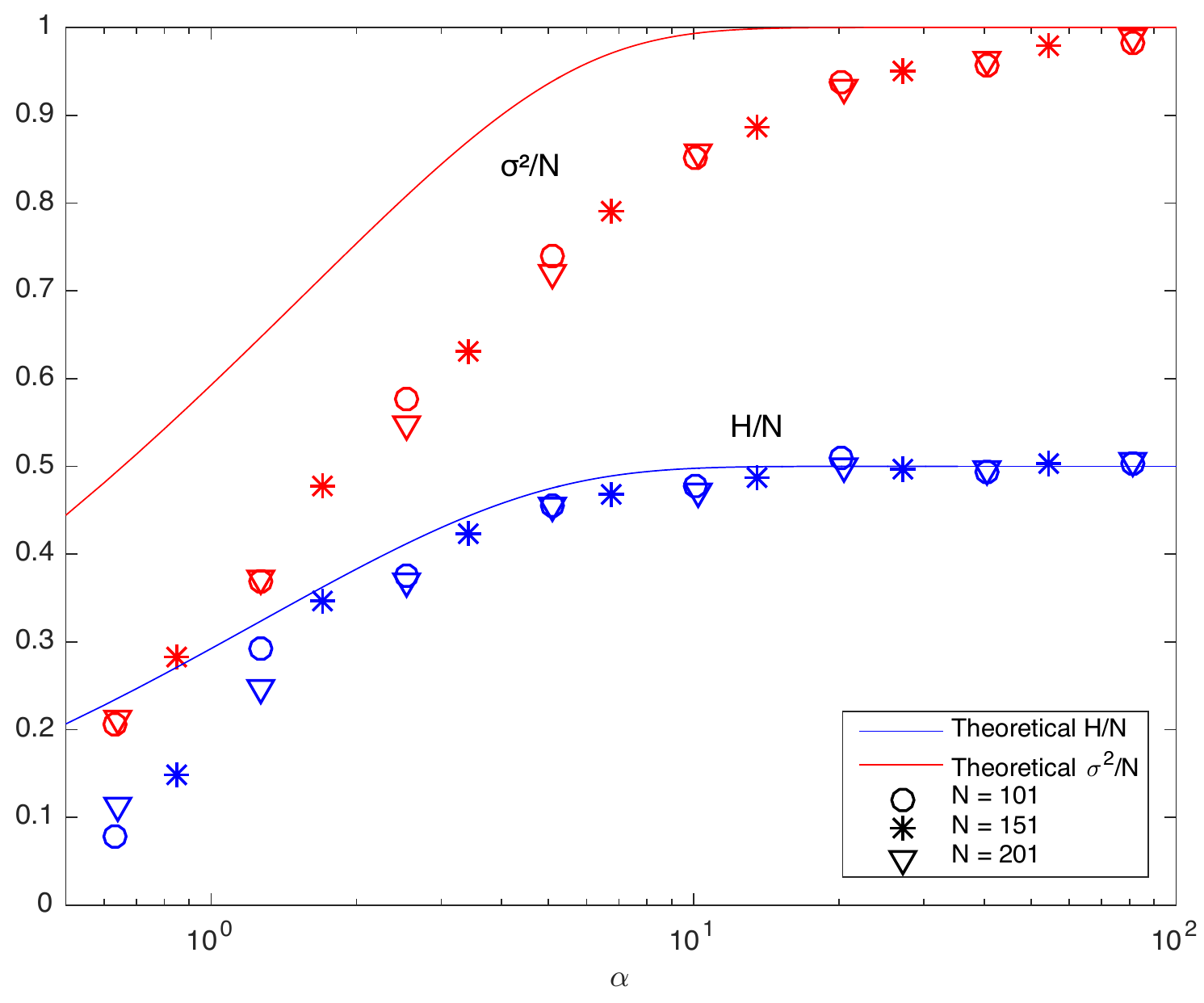}
\caption{\label{sigsqH}
Model and simulation results for $\sigma^2$ and $H$ for the linear payoff game. (Each point is averaged over 10 runs with $10^7$ time steps each).}
\end{figure}

The fluctuations of attendance $\sigma^2=\langle A^2\rangle=H+\varphi N/2$ with $H=\frac{1}{P}\sum_\mu\langle A |\mu\rangle^2=\frac{N}{2}(1-c(\phi_1-\phi_2)))^2$ are compared to simulations in Fig. \ref{sigsqH}. These are clearly significantly overestimated by the model. (Similar results are found for the sign-payoff game and model.) Following the exposition in appendix \ref{Hamiltonian}, the reasons for this discrepancy is quite clear. The model always overestimates the fluctuations $S_t$, and since we are assuming that only the frozen agents contribute to $\langle A|\mu\rangle$ we also miss the contribution of the fickle agents to reduce $H$. There seems to be a quite clear path to improve the model along these lines, which is left for future work. Here we opt for the simplicity of solving the present model and the fact that it does give quantitative agreement with distribution of realtive strategy scores.

As a next step we can find the score distributions by solving the master equation on an integer chain. In contrast to the t game where scores are only updated by 0 or $\pm$1,  we now have to consider longer range hopping where scores are updated by integer steps in the range $-N$ to $N$. Taking into account the individual time averaged step size $\Delta_\pm=\Delta_{\text{fit}}\mp\frac{1}{2}$ (for $x>0$ and $x<0$ respectively) and the fact that $\xi^{\mu(t)} A_t$ has variance $N/2$, we expect that the jump propabilities are well represented by 
a normal distribution (for a jump from $x$ to $x'$) 
\begin{equation}
p_{x\rightarrow x'}={\cal N}_{(x'-x)}(\Delta_\pm,\sqrt{\frac{N}{2}})\,.
\end{equation}
The master equation takes the form
\begin{equation}
P_x(t+1)=\sum_{x'}p_{x'\rightarrow x}P_{x'}(t)\,.
\end{equation}
Taking the continuum limit over space and ignoring complications due to the boundary $x=0$, this can be solved in terms of exponential localization for fickle agents ($\Delta_+<0$ and $\Delta_->0$) 
and diffusion with a drift for frozen agents ($\Delta_+>0$ or $\Delta_-<0$). For fickle agents the score distributions are given by 
\begin{equation}
P(x)\sim e^{\mp 4|\Delta_\pm|x/N}
\end{equation}
for $x>0$ and $x<0$ respectively, which in the large $\alpha$ limit reduces to $P(x)\sim e^{\mp 2x/N}$. For frozen agents the distributions are given by 
\begin{equation}
P(x,t)={\cal N}_{x}(\Delta_{\pm}t,\sqrt{\frac{Nt}{2}})\,,
\end{equation}
for positively and negatively frozen agents respectively.

\end{document}